\documentclass[traditabstract]{aa} 
\usepackage{subfigure}                                   
\usepackage{graphicx}
\usepackage{txfonts}
\begin{document}

   \title{Finding binary active galactic nuclei by the centroid shift in imaging surveys}

   \author{Yuan Liu
          \inst{1}
          }

   \institute{Key Laboratory of Particle Astrophysics, Institute of High
Energy Physics, Chinese Academy of Sciences, P.O.Box 918-3, Beijing
100049, China\\
              \email{liuyuan@ihep.ac.cn}}

   \date{...}


  \abstract
   {The census of binary active galactic nuclei (AGNs) is important in order to understand  the merging history of galaxies and the triggering of AGNs. However, there is still no efficient method for selecting the candidates of binary AGNs. The non-synchronous variations of the two AGNs in one binary system will induce the shift of the image centroid. Since the astrometric error is normally much smaller than the angular resolution of telescopes, it is possible to detect such shifts even in the unresolved system via multi-epoch observations. We perform some simulations and find that hundreds of observations are required to discover compact binary AGNs. This method is suitable for the future large-scale surveys, e.g., the Large Synoptic Survey Telescope, and it might lead to a large sample of binary AGNs with a 1-2 yr survey.}

   {}


   \maketitle
%

\section{Introduction}
In the current framework of the evolution of the cosmological structure, galaxies assemble their mass through mergers and gas accretion (Hernquist 1989; Kauffmann \& Haehnelt 2000; Volonteri et al. 2003; Hopkins et al. 2008). The major merger is believed to be an important  mechanism that triggers the central supermassive black holes to become active galactic nuclei (AGNs). Therefore, binary AGNs should be common in merging galaxies. The fraction of binary AGNs is a key diagnostic of the AGN triggering and gravitational wave searching, though there are still numerous    theoretical and   observational uncertainties (Van Wassenhove et al. 2012; Blecha et al. 2013; Dotti et al. 2015; Ravi et al. 2015). Binary AGNs are also valuable for the investigation of the relation between the growth of black holes and the properties of host galaxies, especially for the pairs at small separation (Foreman et al. 2009; Kormendy \& Ho 2013; Medling et al. 2015). Thus, a large sample of binary AGNs with a separation smaller than kpc scale is desired.

Various techniques have been adopted to search for binary AGNs (Liu et al. 2010; Fabbiano et al. 2011; Tsalmantza et al. 2011; Eracleous et al. 2012; Deane et al. 2014; Graham et al. 2015). It is a low probability event to serendipitously find binary AGNs with the distance smaller than kpc scale. An efficient method is needed to find a collection of candidates of binary AGNs in a large-scale survey, just like the color-color diagram used to find quasars with Sloan Digital Sky Survey (SDSS) data. It has been suspected that double-peaked AGNs, for example,  those that have double-peaked [O III] lines, are  binary AGN candidates. There have been extensive works on the properties of double-peaked AGNs (Wang et al. 2009; Smith et al. 2010; Shen et al. 2011). Although some double-peaked AGNs are indeed binary AGNs, it turns out that  this method is not very efficient. Fu et al. (2012) found that only $\sim$2\% of  double-peaked AGNs are the result of the orbital motion of two AGNs. Most  double-peaked AGNs are driven by the motion of the gas in one AGN. In a reverberation mapping campaign, Barth et al. (2015) also found that the velocity shift of broad lines can mimic the signature of binary AGNs and will limit their detectability in spectroscopy.
Therefore, we still need  an efficient method to search for binary AGNs.

We propose a novel method which is suitable for imaging surveys.
The cost of an imaging survey is much lower than a spectroscopy survey, and there are several ongoing and future large-scale imaging surveys. In Sect. 2, we explain the idea of this method and show its feasibility with some simple simulations. In Sect. 3, we discuss the implications of this method and present our conclusions.


\section{Method and simulation}
\subsection{Illustration of the method}
\begin{figure}[!htp]
 \centering
   \includegraphics[width=0.40\textwidth]{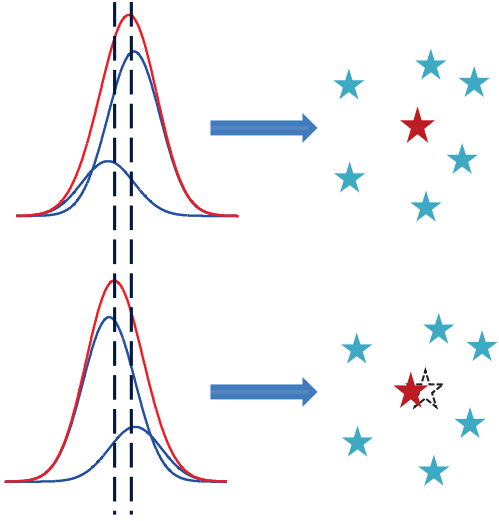}
   \caption{Illustration of the change of a one-dimensional PSF profile (\textit{left}) and image positions (\textit{right}) due to the variation of the two AGNs in one binary system (see the text for details).}
 \label{illu}
 \end{figure}

The idea of the new method is explained in Figure \ref{illu}. If there are two AGNs in the merging galaxies with separation smaller than the angular resolution of telescopes, only a single source can be detected on the image. The image should somewhat deviate from the point-spread function (PSF); however, we cannot identify if it is a binary AGN or an extended source produced in another way, e.g., host galaxies, jets from AGNs, or supernovae. As a by-product of future imaging surveys, the candidates of binary AGNs can be further selected from these extended sources. One of the important features of AGNs is  violent and frequent variation in broad bands. Hence if the same source is observed  many times, the contrast of the luminosities of the two AGNs changes with time (the likely case before the merger of the two accretion disks). As a result, we still cannot resolve the two sources on the image, but the peak or centroid of the source will shift slightly. If the time scale of the orbital evolution of two AGNs is much longer than the duration of the multi-epoch imaging survey, the shift will only occur along the direction of the two AGNs, which is unique compared with other motions, for example,  proper motion, blobs in jet, and other outflow/inflow. \footnote{If the positions of the two AGNs significantly change during the imaging survey, e.g., two AGNs at sub-pc scale with an orbital period of decades, the pattern of the shift could be confused with other mechanisms. In this case, the spectroscopy method is more suitable.} The minimum shift that we can detect is the astrometric error (or positional error) of the instrument, which can be smaller than the angular resolution by one to two orders of magnitude. Therefore, it is possible to identify a binary AGN at a separation much smaller than  the angular resolution of  telescopes. In practice, the centroid of the image will also be randomly distorted by the astrometric error. As a result, the distribution of the centroid should be elongated along the direction of the binary AGN. We will then perform simulations to find the number of observations required to identify this deviation from the astrometric error.

\subsection{Simulations}
To simulate the variation of AGNs, we adopt the long-term light curve of NGC 5548 at 5100 $\AA$ as the template for variations, which is the longest light curve of a single AGN obtained to date\footnote{http://www.astronomy.ohio-state.edu/$\sim$agnwatch/n5548/lcv/c5100.dat}, and list some steps in the simulation process.

1.      The mean luminosities of the two AGNs in one binary system are assumed to be the same as the mean luminosity of the light curve of NGC 5548.

2.      Two luminosities  from the light curve of NGC 5548 are randomly selected as the luminosities of the two AGNs.

3.      Since the centroid only shifts along the direction of the binary AGN, we simply discuss the one-dimensional distribution along this direction. The centroid of the source ${x_c}$  is calculated by  ${x_c} = \frac{{{x_1}{L_1} + {x_2}{L_2}}}{{{L_1} + {L_2}}}$, where $x_1$ and $x_2$ are the positions of the two AGNs, and $L_1$ and $L_2$ are the luminosities selected in Step 2.

4.      The centroid obtained in Step 3 is added with an astrometric error, which is randomly drawn from a Gaussian distribution with the mean $\mu=0$ and $\sigma=$ astrometric error.

5. Steps 2-4 are repeated $N$ times, i.e., this source is observed $N$ times. We assume that the total duration of the campaign is long enough, therefore it is possible to observe every luminosity pair. Then the Kolmogorov-Smirnov (K-S) test is adopted to calculate the significance of the difference between the  distribution of ${x_c}$  from these $N$ observations and the distribution of astrometric error ($\mu=\overline{x_c}$). The difference is thought to be significant if $P_{K-S}<0.01$.

6. Since $N$ is a finite number, the significance from a K-S test changes from sample to sample. Thus, we repeat Steps 2-5 40000 times and build the distribution of the significance from the K-S tests.

7.       To determine the number of observations required to find binary AGNs, $N$ is gradually increased until the difference is significant in 99\% of the sets of observations according to the distribution of the significance from the K-S tests built in Step 6.

\subsection{Results}
We assume the separation between the two AGNs is 0.2 arcsec, which corresponds to 0.37 kpc at z=0.1 (the bulk of the SDSS galaxy sample) assuming a standard cosmology. The mean seeing at the site of Large Synoptic Survey Telescope (LSST) is $\sim$0.67 arcsec  (LSST Science Collaboration et al. 2009). Thus, as discussed in Sect. 2.1, it is possible to find that the source is extended but difficult to identify it as  a binary AGN.
The requirement for astrometry with LSST is $\sim$0.01 arcsec (LSST Science Collaboration et al. 2009).
If the luminosities of the two AGNs are the same, the number of observations required to discover binary AGNs  is 115 according to the procedure in Sect. 2.2. If we change the assumption in Step 1, i.e., the luminosity of one AGN is higher than the other one by a factor of 3, the amplitude of the shift is smaller. However, the distribution of $x_c$ is skewed in this case, which is helpful in order to identify the difference in the distribution of astrometric error. As a result, the number of observations required  increased only  moderately  to 206.
Figures 2 and 3 show the distribution of $x_c$ and $P_{K-S}$ for these two cases. Figure 4  presents the number of observations required as the function of luminosity ratios.

 \begin{figure}[!htp]
 \centering
   \includegraphics[width=0.45\textwidth]{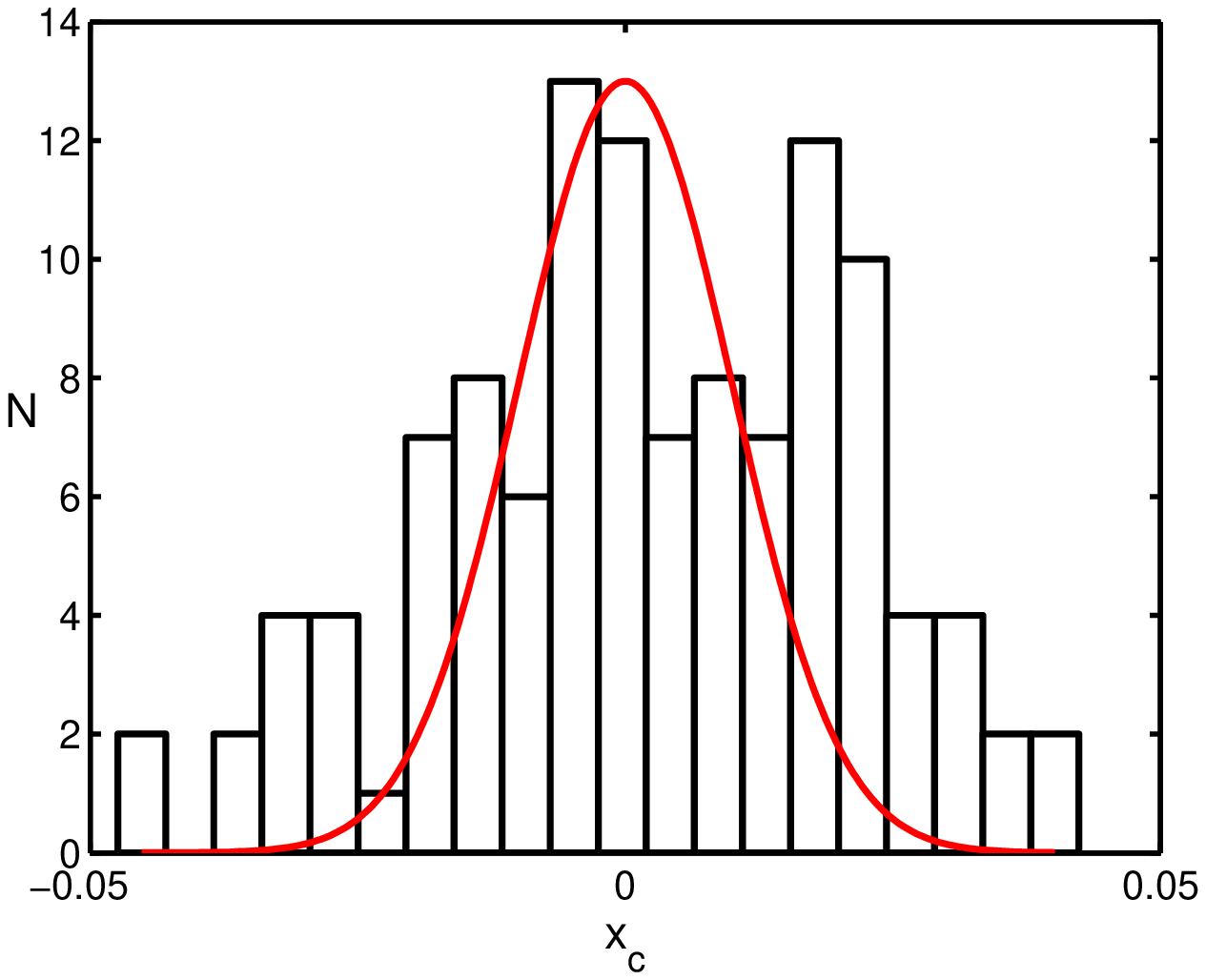}
   \includegraphics[width=0.45\textwidth]{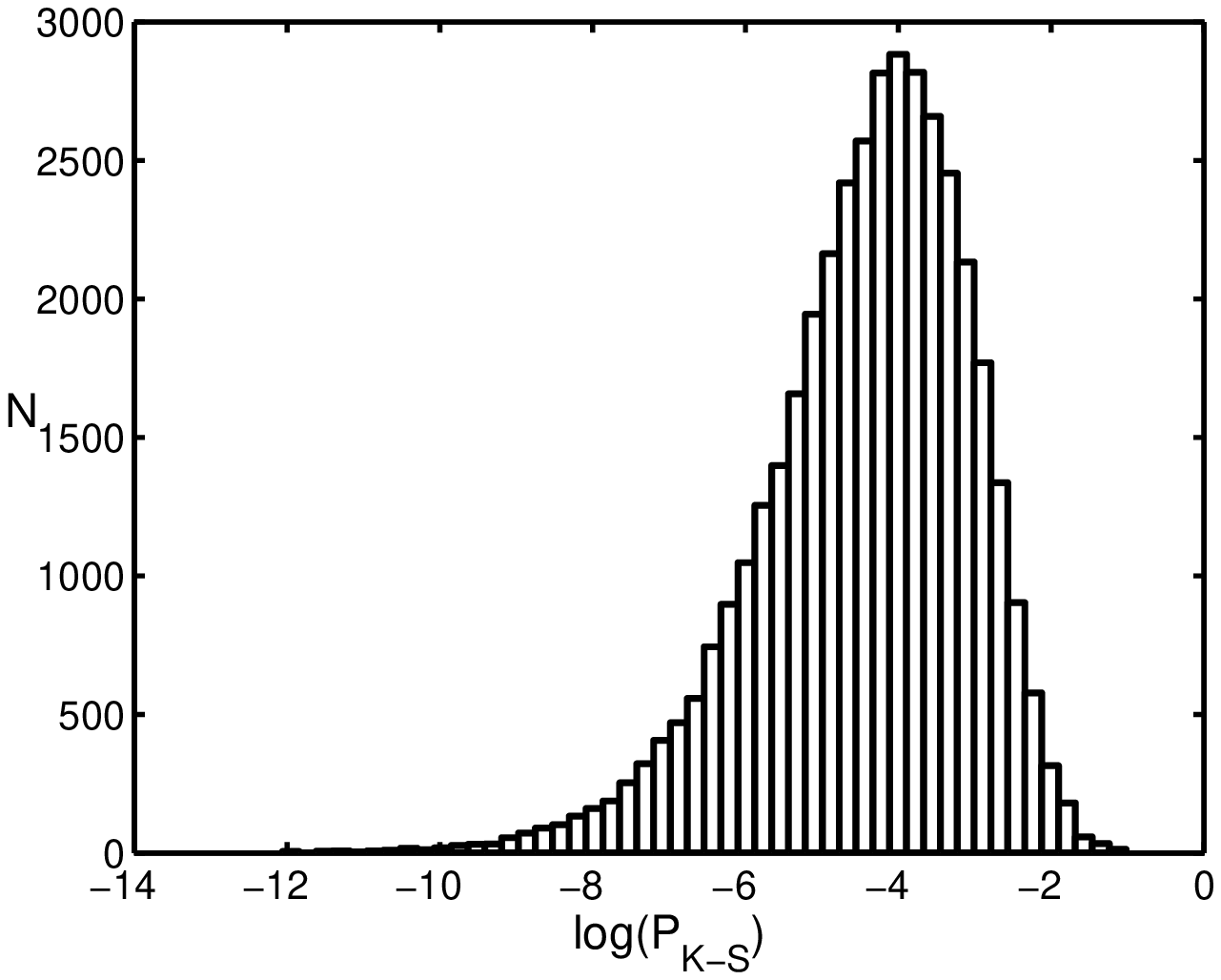}
   \caption{\textit{Top}: histogram of the centroid $x_c$ of one set of 115 observations ($\overline{L_1}=\overline{L_2}$). The solid line is  the distribution of the astrometric error ($\mu=\overline{x_c}$, $\sigma=0.01$ arcsec); \textit{Bottom}: histogram of $P_{K-S}$ of all 40000 sets of observations.}
 \label{d1}
 \end{figure}

 \begin{figure}[!htp]
 \centering
   \includegraphics[width=0.45\textwidth]{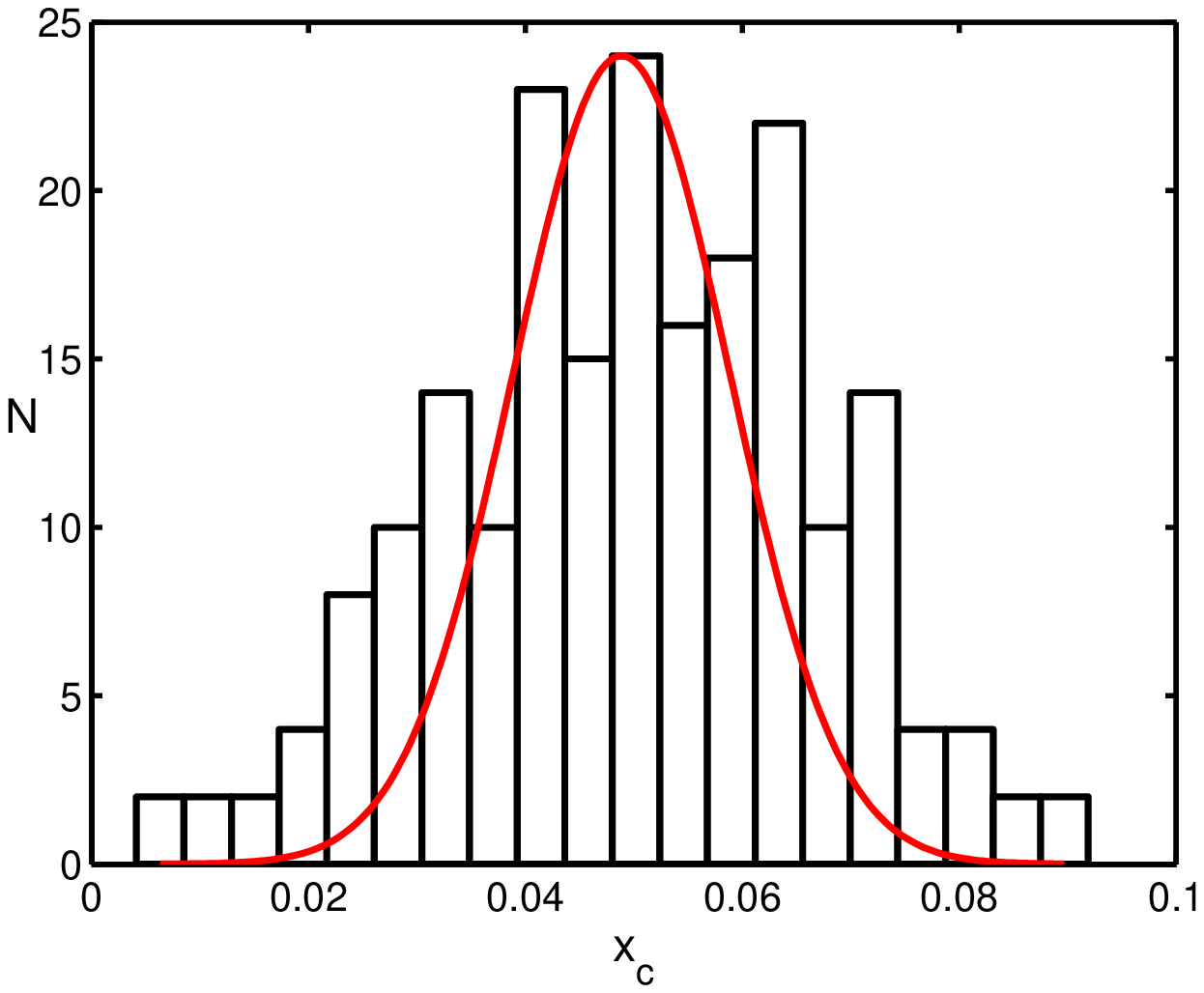}
   \includegraphics[width=0.45\textwidth]{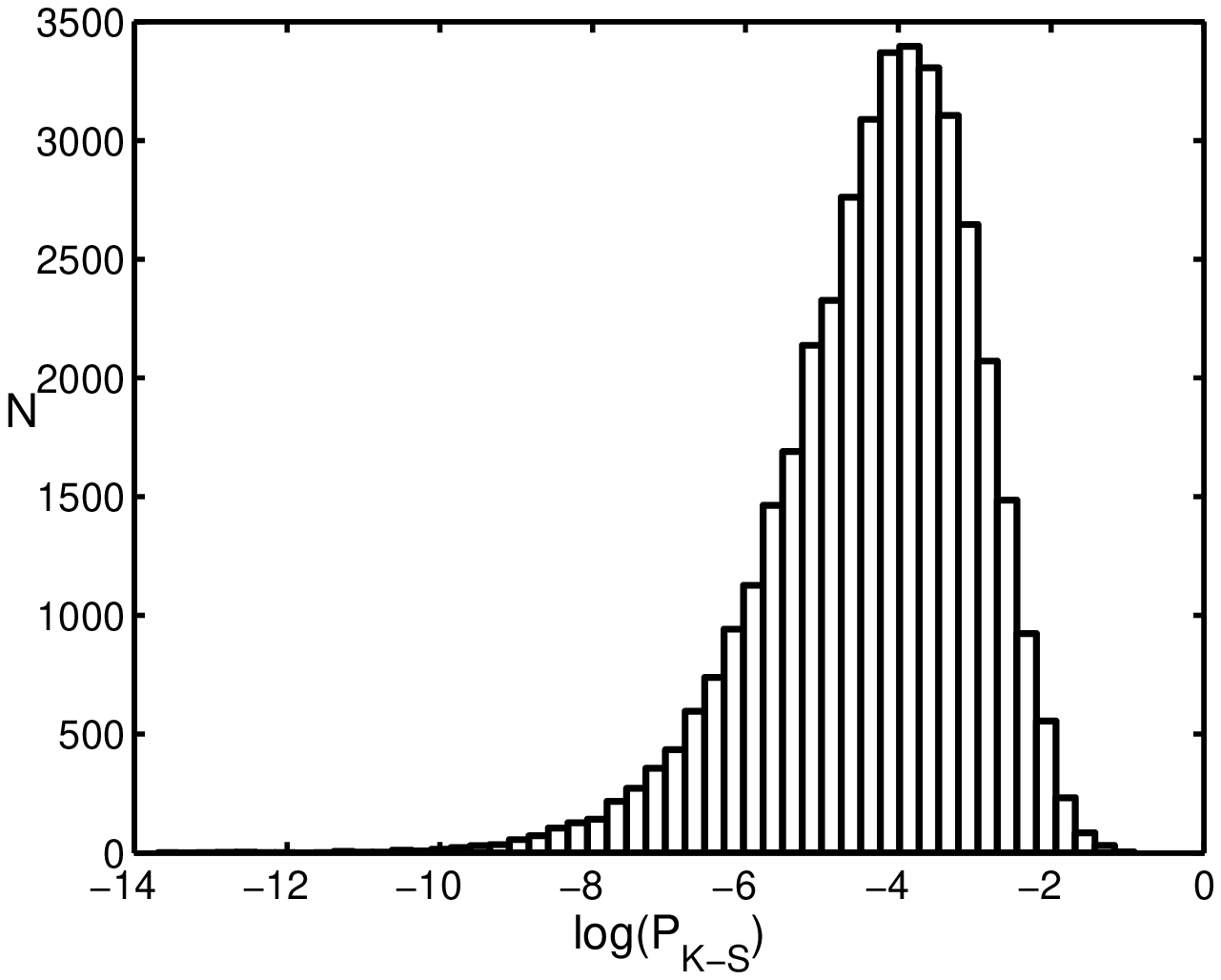}
   \caption{The same as Figure \ref{d1} but for $\overline{L_1}=\overline{L_2}/3$.}
 \label{d2}
 \end{figure}

 \begin{figure}[!htp]
 \centering
   \includegraphics[width=0.45\textwidth]{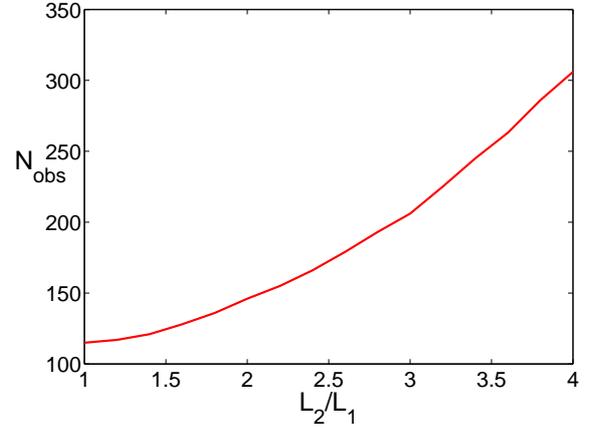}
   \caption{Relation between the luminosity ratio and the number of observations required.}
 \label{d3}
 \end{figure}


\section{Discussions and conclusions}

As was shown in Sect. 2.3, several hundreds of observations are necessary to find binary AGNs with the method proposed here. In principle, this method is suitable for all bands. Although the angular resolution of the current X-ray or radio observation is high enough, it is more appropriate to apply this method to the large-scale survey in optical band, e.g., Pan-STARRS and LSST. The LSST can scan the same area of the sky every 2-3 days; thus a survey of 1-2 years is enough to fulfill the required number of observations.

The astrometric error is affected by the signal-to-noise ratio (S/N) of the targets. Under the bright source limit, the astrometric error will be simply improved by a factor of S/N compared with the seeing (King 1983; Mighell 2005). Thus, for LSST, a signal-to-noise ratio of 100 is high enough to achieve the astrometric error of 0.01 arcsec. According to the exposure calculator of LSST\footnote{http://dls.physics.ucdavis.edu:8080/etc4\_3work/servlets/LsstEtc.html}, the source brighter than 20.5 mag can fulfill this requirement using one exposure (15 sec). The final astrometric error is limited by other systematic errors, e.g., instrumental distortion, atmosphere differential refraction, and atmosphere turbulence. Nevertheless, the enhancement of 100-200 compared with the seeing is achievable by current techniques (Fritz et al. 2010; Trippe et al. 2010).

In practical observations, the astrometric error distribution can be constructed from other reference stars. Since  the direction of the two AGNs is not known, the centroid can be projected to an arbitrary direction at first. Then we rotate the initial direction to find a direction with the maximum variation of the centroid as the possible direction of the two AGNs. Finally, the distribution of the centroid in this direction is compared to the distribution of the astrometric error. There should be no significant difference between the distribution of the centroid in the perpendicular direction and the distribution of the astrometric error, otherwise this source should not be a binary AGN.

Combining the temporal behavior of the shift, it is easy to distinguish the random shift due to a binary AGN from the shift produced by proper motion, a fast moving jet, a comet, or an asteroid. The stochastic variation of AGNs is also different from the outburst or periodic variation, e.g., Cepheids, novae, and supernovae. However, a slow moving jet or the light reflection from a single AGN can mimic the centroid shift due to binary AGNs. To confirm the candidates found by the shift in the image, we can perform follow-up high resolution imaging (AO or interferometry), optical spectroscopy,  and X-ray observations.

The imaging method is more sensitive to separation that is significantly larger than the astrometric error; the spectroscopy method can still work at  separations smaller than the astrometric error if the dynamical imprint of the two black holes is evident. However, the imaging method can  nicely complement the spectroscopy method for the binary AGNs with a face-on orbital plane (the velocity in the line of sight is too small to find the shift of the lines). The imaging method is limited to the case where the continua are dominated by AGNs, i.e., two type 1 AGNs. The contamination from the host galaxy will also suppress the amplitude of the shift. If the signal-to-noise ratio is high enough, this method can be applied to high redshift. There is a peak of the angular diameter distance with the redshift in $\Lambda$CDM cosmology, of which the angular diameter corresponds to a scale of about 9 kpc/arcsec. Thus, if the astrometric error is smaller than 0.1 arcsec, we can discover binary AGNs with a distance of about 1 kpc at all redshifts.

\begin{acknowledgements}
The author thanks the referee for useful comments that
improved the paper. This work is supported by the National Natural Science Foundation of
China under grant Nos. 11103019, 11133002, and 11103022.
\end{acknowledgements}

\end{document}